\documentclass{eas}
\usepackage{natbib,aas_macros,epsfig}
\bibpunct{(}{)}{;}{a}{}{,}
\def\spose#1{\hbox to 0pt{#1\hss}}

\def\ledd{${\rm L}_{\rm E}$}

\def\lta{\mathrel{\spose{\lower 3pt\hbox{$\mathchar"218$}}
     \raise 2.0pt\hbox{$\mathchar"13C$}}}
\def\gta{\mathrel{\spose{\lower 3pt\hbox{$\mathchar"218$}}
     \raise 2.0pt\hbox{$\mathchar"13E$}}}
\def\pd#1#2{{\partial #1 \over \partial #2}}

\def\fd#1#2{{{\rm d} #1 \over {\rm d} #2}}

\begin{document}

\title{Thin discs, thick discs and transition zones}

\author{Guillaume Dubus}\address{California Institute of Technology, MC 130-33, Pasadena, CA 91125 {\tt gd@tapir.caltech.edu}}%

\begin{abstract}
Accretion onto a compact object must occur through a disc when the
material has some initial angular momentum. Thin discs and the thicker
{\em low radiative efficiency accretion flows} (LRAFs) are solutions
to this problem that have been widely studied and applied. This is an
introduction to these accretion flows within the context of X-ray
binaries and cataclysmic variables.
\end{abstract}

\maketitle

Accretion describes the increase of the mass of an astrophysical
object when matter is deposited onto it. The conversion of a fraction
of the gravitational potential energy of the accreted material into
radiation can dominate the emission from the system. Accretion is a
major source of energy in a varied lot of astrophysical objects
including interacting close binaries, active galactic nuclei and
protostellar systems (the standard textbook is \citealt{fkr}).

My focus is on close binaries in which a normal star transfers mass
onto a black hole/neutron star (X-ray binaries, XRBs) or white dwarf
(cataclysmic variables, CVs) via Roche-lobe overflow. Exhaustive
reviews of their observational properties may be found in \cite{xrbs}
and \cite{cvs} respectively. In these systems, an accretion disc forms
around the compact object due to angular momentum conservation. The
detailed characteristics of the flow depend upon how matter dissipates
its potential energy and gets rid of its angular momentum in the
disc. This basically leads to two extremes: the ``low radiative
efficiency'' accretion flows (LRAFs, which are geometrically thick)
and the radiatively efficient, geometrically thin discs (hence the
title). This is an introduction to these solutions. \S1 groups the
basic tools needed to present the solutions in \S2 (thin discs) and
\S3 (thick discs); \S4 is a brief discussion of the transition from
one to the other.

\section{The building blocks}

\subsection{Accretion efficiency, Eddington luminosity}
In Newtonian dynamics, the available gravitational energy from a small
mass $m$ moved from infinity to the surface $R_\star$ of the accretor
is $G M_\star m / R_\star$. The energy released can be a sizeable
fraction $\eta$ of the rest-mass energy $m c^2$ when the accretor is a
neutron star or black hole: assuming a thin disc and free fall at the
innermost stable circular orbit then $\eta\approx 0.42$ for a
maximally rotating Kerr black hole (\citealt{nt}; see also
\citealt{gammie}).

The Eddington luminosity \ledd{} is the luminosity for which radiation pressure exactly balances the gravitational pull on the accreted material. Accretion is quenched above this luminosity \citep[but see][]{begelman,shaviv}. For a spherical inflow of ionised hydrogen such that the radiation pressure is due to Thomson scattering on electrons and the dominant gravitational pull is on the protons, \ledd{} is:
\begin{equation}
\frac{GM_\star {\rm m}_{\rm p}}{R^2}= \frac{{\rm L}_{\rm E}}{4\pi R^2} \frac{\sigma_{\rm T}}{c} {\rm ~~hence~~} {\rm L}_{\rm E}\approx 10^{38} ~(M_\star/{\rm M}_\odot) {\rm ~erg}\cdot{\rm s}^{-1}
\end{equation}
The luminosities of XRBs and CVs are generally lower than this limit \citep{xrbs,cvs}. The Eddington mass accretion rate $\dot{\rm M}_{\rm E}$ is defined as:
\begin{equation}
\eta\dot{\rm M}_{\rm E}c^2={\rm L}_{\rm E} {\rm ~~i.e.~~} \dot{\rm M}_{\rm E}\approx 10^{18} ~(M_\star/{\rm M}_\odot)(0.1/\eta) {\rm ~g~s}^{-1}
\end{equation}
 
\subsection{Accretion flow temperature}
Assuming the gravitational energy is released into radiation at the surface of the compact object, the minimum temperature of the flow is that of the black body radiating the same luminosity. At the Eddington limit this gives:
\begin{equation}
4\pi R^2_\star \sigma T^4_{\rm bb}={\rm L}_{\rm E} {\rm ~~i.e.~~} {\rm k}T_{\rm bb}\approx 1.5 ~{\rm keV}~({\rm L}_{\rm E}/10^{38}{\rm erg~s}^{-1})^{1/4} (R_\star/ 10{\rm ~km})^{-1/2}
\end{equation}
Accreting black holes and neutron stars should radiate mainly in soft X-rays while white dwarfs ($R_\star\approx 10^4$~km) should radiate in UV. This is consistent with observations \citep{xrbs,cvs}.

On the other hand, the maximum temperature is obtained when all the
gravitational potential energy is transformed into thermal energy $e$
without radiation losses (adiabatic flow):
\begin{equation} e=\frac{3}{2}\frac{{\rm k}T_{\rm g}}{\mu m_{\rm H}}= \frac{G
M_\star}{R_\star} {\rm ~~i.e.~~} {\rm k}T_{\rm g}\approx 45 {\rm
~MeV~}(M_\star/{\rm M}_\odot) (R_\star/ 10{\rm ~km})^{-1} 
\label{tg}
\end{equation}
assuming ionised hydrogen ($\mu=0.5$). This is twice the virial
temperature of bound particles in circular orbit at
$R_{\star}$. Accretion onto a compact object can power the emission
of gamma rays.
 
\subsection{Disc formation}
Matter infalling onto the compact object will generally have some
non-zero angular momentum. A particle with a ballistic trajectory will
therefore have $R^2\Omega=(R^2\Omega)_{{\rm t}=0}$. At the radius of
closest approach to the compact object $1/2 (R \Omega)_{\rm c}^2 =
GM/R_{\rm c}$ so that $2GMR_{\rm c}=(R^2\Omega)_0^2$. Only a particle
with very low angular momentum can directly hit the compact object
($R_{\rm c}<R_\star$). This condition is not met in compact binaries
where matter comes from an orbiting companion and the stream will go
round the compact star and intersect itself. Subsequent shocks lead to
the dispersion of energy and the stream settles onto a circular orbit
with the initial angular momentum (see \citealt{lubow} for details).

A steady supply of matter with some specific initial angular momentum
piles up in a ring at this circularisation radius. There is no
accretion unless some matter can transfer its angular momentum to
reach smaller orbits. Under such a process, an accretion disc forms
extending down to the compact object.

\subsection{Angular momentum transport}

One process by which particles may exchange angular momentum is
viscosity (see e.g. \citealt{terquem} for an introduction). In gas
kinetic theory, molecular viscosity arises from the exchange of
momentum across the surface of a fluid element. The resulting force is
proportional to $\nu_{\rm mol}\sim \lambda u$ where $\lambda$ is the
mean free path between collisions and $u$ is the mean thermal speed of
the particles. Assuming particles in a plasma interacting only via
Coulomb forces leads to:
\begin{eqnarray*}
& \lambda \sim & 1~{\rm cm}~(T/10^5~{\rm K})~(\rho/10^{-8}~{\rm g~cm}^{-3}) \nonumber \\
 & u \sim & 10^6{\rm ~cm~s}^{-1}~(T/10^5{\rm ~K})^{1/2}
\end{eqnarray*}
using typical accretion disc temperatures and densities
\citep{fkr}. The viscosity coefficient is therefore $\nu_{\rm mol}\sim
10^6{\rm ~cm^2~s^{-1}}$.

Anticipating a little bit on result from the thin disc model, we can
get an order-of-magnitude observational estimate for viscosity in
accretion discs. If the eruptions of dwarf novae, a sub-class of CVs
(see \S\ref{cv}), are due to the accretion of matter coming from the
outer regions of a disc and transported by viscosity then $\nu_{\rm
disc}\sim R_{\rm d} v_{\rm r}$ where $R_{\rm d}$ is the outer disc
radius and $v_{\rm r}$ is the radial velocity of the infalling
matter. $v_{\rm r}\sim R_{\rm d}/\tau_{\rm e}$ where $\tau_{\rm e}$ is
the timescale of the eruption. The disc radius can be estimated using
e.g. the circularisation radius (\S1.3) and typically will be $\sim
10^{10}$~cm. For an eruption lasting a day, $\nu_{\rm disc}\sim
10^{15}{\rm ~cm^2~s^{-1}}$ i.e. $\nu_{\rm disc}\gg \nu_{\rm mol}$.
Molecular viscosity is much too weak to account for accretion in
discs.

The process by which angular momentum is transported has been the
subject of intense research with turbulent transport the prime
suspect. Turbulent viscosity is often modelled using the Navier-Stokes
formalism but with a coefficient $\nu_{\rm turb}\sim \lambda_{\rm
turb} u_{\rm turb}$ where the scale and speed are those of the
turbulent eddies. These are dynamic properties of the fluid, not
intrinsic as with molecular viscosity. In a disc, the size and speed
of the eddies are likely to be limited by the scale-height $H$ and
sound speed $c_{\rm s}$:
\begin{equation}
\nu=\alpha c_{\rm s} H \label{ssp}
\end{equation}
where $\alpha$ is a parameter $<1$. This is the famous $\alpha$
parameterisation of disc viscosity first described in a landmark paper
by \cite{ss}. In the following I will use this parameterisation.

One would rather want to derive $\nu$ from first principles. At
present, only turbulence arising from the magneto-rotational
instability (MRI) has succeeded in predicting any significant
viscosity (see \citealt{bh} for a review). Fortunately, there is some
sense in modelling the angular momentum transport and energy
dissipation of MHD turbulence using the $\alpha$ parameterisation
\citep{bp}.  However, the mechanism may not work in weakly ionised
flows and this is a problem for models with a cold accretion disc
\citep{stepinski,menou}. Putative global hydrodynamical instabilities
might take over in these conditions. There are other possibilities for
angular momentum transport including instabilities in a
self-gravitating disc (e.g. \citealt{bp} and references therein),
spiral waves excited by tidal torques \citep{spruit}, hydromagnetic
winds launched from the disc \citep{blapay}, radiative viscosity
\citep[e.g.][]{loeb} etc. However, these only apply in specific
conditions.

\subsection{Vertically integrated disc equations} 
The disc equations derive from the equations of fluid dynamics
combined with a model for viscosity (which contains the turbulent
magnetic field contributions in the MRI case), an equation of state
for the gas and a description of the radiative processes. A standard set of assumptions to start with is:
\begin{itemize}
\item Axisymetry so that $\partial/\partial\phi=0$ in cylindrical coordinates.
\item The only non-zero component of the stress tensor is the azimuthal shear $\tau_{r\phi}$.
\item The gas is perfect so the internal energy per unit mass is $e=c_vT$ and the gas pressure $P=\rho e (\gamma-1)$ with $\gamma=c_p/c_v$. Radiation pressure is neglected here for simplicity.
\end{itemize}
In addition I suppose the disc self-gravity is negligible (this
assumption must be dropped for protostellar discs and AGNs) and that
relativistic corrections are negligible (which is fine when more than
a few gravitational radii away from the compact object). Assuming
hydrostatic balance so that $v_{\rm z}=0$, the vertical momentum
conservation is ($P=\rho c_s^2$):
\begin{equation}
\pd{P}{z} = -\rho g_z {\rm ~~i.e.~~}
\pd{\ln P}{\ln z}=-\frac{\Omega^2_{\rm K}}{c_s^2}~ z^2 \left[1+\frac{z^2}{R^2}\right]^{-\frac{3}{2}} \label{hydro}
\end{equation}
If the height $H$ from the midplane of the disc is $\lta R$ high order terms in $z/R$ can be neglected. This equation can be integrated analytically for a perfect gas yielding $P(z)$ and $\rho(z)$. Averaging $P$ over $z$ gives a relationship between $H$ and the mid-plane sound speed which, to a factor of order unity, is :
\begin{equation}
H= c_s/\Omega_{\rm K}
\end{equation}
The scale-height $c_s/\Omega_{\rm K}$ appears in Eq.~\ref{hydro}. The detailed vertical balance can be bypassed by assuming the disc height is given by this relationship. The Shakura-Sunyaev prescription for the viscosity becomes
\begin{equation}
\nu=\alpha c_s^2 / \Omega_{\rm K}
\end{equation}
Note that $\nu$ is effectively integrated over $z$. With a Navier-Stokes formulation of the viscosity, the integrated stress is:
$$\tau_{r\phi}=\nu \Sigma R \pd{\Omega}{R}$$
where $\Sigma=2 \rho_o H$ is the column density and $\rho_o$ is the mean density. The radial evolution equations are then integrated and decoupled from $z$, resulting in a set of time-dependent 1D equations:
\begin{eqnarray}
\pd{\Sigma}{t}+\frac{1}{R}\pd{}{R}(\Sigma R v_r) & = & 0 \label{mass}\\
\pd{v_r}{t}+v_r \pd{v_r}{R} & = & R\Omega^2-R\Omega_{\rm K}^2-\frac{1}{\rho_o}\pd{P}{R} 
\label{radial}\\
\pd{R^2 \Omega}{t}+v_r\pd{R^2 \Omega}{R} & = & \frac{1}{\Sigma
R}\pd{}{R}\left(R^2\tau_{r\phi}\right)  = \frac{1}{\Sigma R}\pd{}{R}\left(\nu 
\Sigma R^3\pd{\Omega}{R}\right) \label{angular}\\ 
T\pd{s}{t}+v_r T\pd{s}{R} & = & \frac{1}{\Sigma}(Q^+_{\rm vis}-Q^-_{\rm rad}) \label{energy}
\end{eqnarray}
In order, we have the mass, radial momentum, angular momentum, and energy conservation equations. $R\Omega$ and $v_r$ are the angular and 
radial velocities while $P=\rho_o c_s^2$. $c_s$ and therefore $T$ are the mid-plane values of the sound speed and temperature in this framework (Eq.~1.8). Again, only the dominant term in the radial component of gravity is kept $g_r\approx R\Omega_{\rm K}$. In the thermal equation, $s$ is the entropy and  $Q^+_{\rm vis}$ is the energy generated by the work of viscous forces in the annulus:
\begin{equation}
Q^+_{\rm vis}=\nu \Sigma \left(R\pd{\Omega}{R}\right)^2
\label{qplus}
\end{equation}
$Q^-_{\rm rad}$ represents radiative losses.  Using
$ds=de+Pd(1/\rho_o)$ the thermal equation can be rewritten as:
\begin{equation}
\pd{e}{t}+v_r \pd{e}{R} = -\frac{P}{\rho_o^2}\left[ \pd{\rho_o}{t}+v_r \pd{\rho_o}{R}\right] +\frac{1}{\Sigma} (Q^+_{\rm vis}-Q^{-}_{\rm rad})
\end{equation}
and $e$ can be written using $T_{\rm g}$ (see Eq.~\ref{tg}):
\begin{equation}
e=\frac{c_s^2}{(\gamma-1)}=\frac{T}{T_{\rm g}} R^2 \Omega_{\rm K}^2
\label{internal}
\end{equation}
Other versions of these equations accommodate additional energy
sources and sinks, conduction or convection terms, radiation pressure,
different proton and electron temperatures, magnetic fields,
relativistic corrections etc.


\subsection{Steady state disc}
A disc in steady state has $\pd{}{t}=0$.
Combining Eqs.~\ref{mass} and \ref{angular} and integrating over $R$ gives:
\begin{equation}
R^2\Omega=\frac{\nu R^2}{v_r}\pd{\Omega}{R}+{\rm ~constant} 
\label{int1}
\end{equation}
Assuming there are no viscous torques at the surface of the compact 
object, the integration constant is $R_\star^2 \Omega_\star$. This
term is very small compared to $R^2\Omega$ far from the boundary layer. Writing $b=1-(R^2_\star \Omega_\star/R^2 \Omega)$, Eq.~\ref{int1} then gives
\begin{equation}
b v_r= \frac{\nu}{R}\pd{\ln \Omega}{\ln R}\label{vr}
\end{equation}
The radial momentum equation can be rewritten using 
Eqs.~\ref{vr} and \ref{internal} as:
\begin{equation}
\frac{\Omega^2}{\Omega^2_{\rm K}}-1=(\gamma-1)\left( 
\frac{T}{T_{\rm 
g}} \right) \pd{\ln P}{\ln R}+(\gamma-1)^2\left( \frac{\alpha}{b} \frac{T}{T_{\rm g}} \right)^2 
\left( \pd{\ln\Omega}{\ln R} \right)^2  \pd{\ln v_{r}}{\ln R}
\label{radialadim}
\end{equation}
Defining the cooling efficiency $f$ as $1-Q^-_{\rm rad}/Q^+_{\rm vis}$, the energy equation becomes:
\begin{equation}
f b \left(\frac{\Omega}{\Omega_{\rm K}} \right)^2 \pd{\ln \Omega}{\ln 
R}=\left( \frac{T}{T_{\rm g}} \right) \left[ \pd{\ln e}{\ln R}-(\gamma-1)\pd{\ln \rho_o}{\ln R} \right] \label{eneradim}
\end{equation}
In a solution to the equations, $f$ should be calculated self-consistently from the radiative processes.  I find this set of dimensionless equations useful when discussing the assumptions that go into a thin or thick disc. 

\section{Radiatively efficient flows: thin discs}

Thin accretion discs are flows in which the energy liberated by accretion is efficiently radiated locally. The next section (\S3) deals with the LRAFs where this is not the case. Thin discs were first described by \cite{pr72} and \cite{ss}. The standard presentation of this model may be found in the review by \citet[][ see also \citealt{fkr}]{pringle}.
	
\subsection{Main properties in steady state}
For a cool steady disc such that $T/T_{\rm g}\ll 1$ then we must have $f\approx 0$ i.e. $Q^+_{\rm vis}\approx Q^-_{\rm rad}$, viscous heating is balanced locally by radiative losses. In other words, the disc is radiatively efficient. The definition Eq.~\ref{internal} of $e$ as a function of $T/T_{\rm g}$ shows that $c_s\ll R\Omega_{\rm K}$ hence $$H/R \ll 1$$ A cool accretion flow is geometrically thin. It is easy to see from the dimensionless equations that $T/T_{\rm g}\ll 1$ further implies that a steady flow is Keplerian with $\Omega\approx \Omega_{\rm K}$, azimuthally supersonic with $v_\phi\approx R\Omega_{\rm K} \gg c_s$ but that the radial inflow velocity 
\begin{equation}
b v_r= -\frac{3}{2} \frac{\nu}{R} \ll c_s \label{vr2}
\end{equation}
is highly subsonic. The mass accretion rate in the disc, defined as
\begin{equation}
\dot{M}=-2\pi R \Sigma v_r \label{mdot}
\end{equation}
is a constant (Eq.~\ref{mass}). The balance of heating and cooling gives:
\begin{equation}
Q^+=\frac{9}{4}\nu \Sigma \Omega_{\rm K}^2 = Q^-= 2 \sigma T_{\rm eff}^4
\label{localbalance}
\end{equation}
The factor 2 accounts for radiation losses from both sides of the disc. This can be rewritten using Eqs.~\ref{vr2}-\ref{mdot} as
\begin{equation}
\sigma T_{\rm eff}^4 = \frac{3b}{8\pi} \frac{GM_\star\dot{M}}{R^3} 
\label{teff}
\end{equation}
The disc dissipates an energy $3/2 G M\dot{M}/R^2 dR$ between $R$ and $R+dR$ far from the inner boundary, three times more than the local release of gravitational energy. The extra energy comes from smaller radii with viscous transport redistributing the way the available gravitational energy is radiated in the thin disc. This equation also shows clearly that the emission of a {\em steady state} disc is independent of the particular heating mechanism (viscosity in this case). Thus, there is no need to know the local angular momentum transport mechanism in order to predict the disc luminosity.

Eclipse mapping has been used to test the radial dependence of the emission. In a high inclination system, the companion gradually eclipses different parts of the disc during its orbital motion. The resulting eclipse profile can be used to map the radial distribution of the disc brightness temperature. Results show that dwarf novae in outburst and persistent CVs (novae-like systems) have temperatures varying as $R^{-3/4}$ in agreement with expectations. On the other hand, quiescent dwarf novae show flat radial temperature distributions and that supports the disc instability model which is discussed later in \S\ref{dim} \citep[not all observations fit this simple picture, see e.g.][]{smak94,cvs,baptista}.

\subsection{Vertical structure \label{vertstruct}}

The thermal balance (Eq.~\ref{localbalance}) is written without any
knowledge of the radiative processes that cool the disc. This is needed
to relate $Q^-_{\rm rad}$ to the quantities appearing in the radial
equations. For instance, one can solve for $Q^-$ by integrating the
hydrostatic balance (Eq.~\ref{hydro}) and vertical radiative transfer
equation (Eq.~\ref{energy}) explicitly. For optically thick
radiation, this is like solving a stellar atmosphere \citep[see][ and
references therein]{hubeny}. The caveat is that additional assumptions
have to be made on how viscous dissipation occurs locally in the
layer (typically $q^+_{\rm vis}\propto P$).

The simple one zone model for the vertical structure has $H\Omega_{\rm
K}=c_s$, $\Sigma=2\rho_0 H$, $\tau =\kappa \Sigma$ and $P= \rho_0 k
T/\mu m_H$. The radiation pressure is neglected and $\kappa$ is the
Kramers opacity (Rosseland mean of bb, bf \& ff processes). If the
layer is optically thick, i.e. the disc cools at the maximum possible
rate, the Eddington approximation for the radiative flux relates
$T_{\rm eff}$ to $T$. These assumptions can be shown retrospectively
to be valid for the range of densities and temperatures considered.
\begin{equation}
Q^-=\frac{4\sigma}{3\tau}T^4 =Q^+=\frac{3}{8\pi}\frac{GM_\star \dot{M}}{R^3}
\label{thick}
\end{equation}
This completes the set of equations and a steady solution can now be found \citep{fkr}:
\begin{eqnarray}
\Sigma&=&5.2~\alpha^{-4/5} \dot{M}_{16}^{7/10} M_1^{1/4} R_{10}^{-3/4}{\rm ~g~cm}^{-2} \nonumber\\
H/R&=&0.02 ~\alpha^{-1/10} \dot{M}_{16}^{3/20} M_1^{-3/8} R_{10}^{1/8}\nonumber\\
T&=&14000 ~\alpha^{-1/5} \dot{M}_{16}^{3/10} M_1^{-1/4} R_{10}^{-3/4} {\rm ~K} \nonumber\\
\tau&=&190 ~\alpha^{-4/5} \dot{M}_{16}^{1/5} \nonumber\\
\nu&=&2\cdot 10^{14}  ~\alpha^{4/5} \dot{M}_{16}^{3/10} M_1^{-1/4} R_{10}^{3/4} {\rm ~cm}^2{\rm~s}^{-1} \nonumber\\
v_r&=&3\cdot 10^{4}  ~\alpha^{4/5} \dot{M}_{16}^{3/10} M_1^{-1/4} R_{10}^{-1/4} {\rm ~cm~s}^{-1} \nonumber
\end{eqnarray}
where $M_1$ is the mass of the compact object in solar units, $R_{10}$ is the radius in units of $10^{10}$ cm and $\dot{M}_{16}$ is the accretion rate in units of $10^{16}$~g~s$^{-1}$. Around a black hole or neutron star the disc can extend down to small radii. As the temperature increases, electron scattering takes over from free-free as the major source of opacity. At very small radii and high accretion rates, radiation pressure can become dominant but the disc is then unstable (see \S\ref{stab}). 

\subsection{Radiation spectrum}

The outgoing spectrum of the optically-thick model is calculated by considering each disc annulus radiates as a blackbody at a temperature $T_{\rm eff}$ and integrating over $R$. The resulting spectrum is a modified disc blackbody \citep[Fig.~20 of][]{fkr}. The maximum temperature of the radiation is reached close to the surface of the accreting object at $R=(7/6)^2 R_\star$ (Eq.~\ref{teff}). $\sigma T_{\rm eff}^4$ is of the order of 1~eV in the case of white dwarfs, 10~eV for massive black holes and 1~keV for neutron stars and stellar mass black holes. The big blue bump in AGNs and the soft X-ray emission in X-ray binaries
 can be interpreted as modified blackbodies from a thin disc. Optical observations leave little doubt that thin discs are present in compact binaries (e.g. double-peaked emission lines, radial temperature profiles, statistics of eclipsing systems, orbital lightcurves etc).

But it was quickly recognized that these objects also show emission at higher energies, which cannot be explained by a thin disc \citep[e.g.][]{ls1975}. One possible origin for this radiation is the boundary layer where the angular velocity of matter decreases rapidly to match that of the compact object. Integrating Eq.~\ref{teff} over $R$ shows only half of the available energy $GM_\star\dot{M}/R_\star^2$ is radiated within the thin disc and the balance is dissipated in the boundary layer (when the object has a surface). Modelling the contribution of the boundary layer to the spectrum of CVs and neutron star XRBs is a difficult problem \citep[see][ and references therein]{regev,popham}. Other likely possibilities include contributions from a tenuous hot corona above the disc (somewhat analogous to the solar corona; see \citealt{poutanen} and references therein) and/or emission from a different type of accretion flow closer to the compact object (see \S\ref{stab} and 3.2) and/or a jet \citep[e.g.][]{mf}.

\subsection{Evolving thin disc: timescales \label{time}}

A cool evolving disc is not necessarily Keplerian and this assumption must be made when studying time-dependent thin discs. Numerical simulations show this is a very good approximation \citep[e.g.][]{ludwig}. The radial momentum equation can be put in a dimensionless form:
\begin{equation}
\frac{v_r^2}{R^2\Omega_{\rm K}^2} \left( \frac{R}{v_r}\pd{\ln v_r}{t}+\pd{\ln v_r}{\ln R}\right) = \frac{\Omega^2}{\Omega^2_{\rm K}}-1-(\gamma-1)\left(\frac{T}{T_{\rm g}} \right) \pd{\ln P}{\ln R} \approx 0
\end{equation}
showing $v_r \ll R\Omega_{\rm K}$, just like the steady case. The conservation of angular momentum with $\Omega=\Omega_{\rm K}$ leads to $v_r\sim \nu/R$ (see Eq.~\ref{vr}).

The major timescales of a thin disc are directly identified from the
dimensionless forms of the other Eqs.~\ref{mass}-\ref{energy}. The
dynamical timescale is:
\begin{equation}
t_{\rm dyn} \sim {R \over v_{\phi}} \sim {1 \over \Omega_{\rm K}} \approx 100{\rm ~s~~} M_1^{-1/2} R_{10}^{3/2}
\end{equation}
This is the shortest timescale over which azimuthal inhomogeneities (e.g. flares) would be expected to vary. Assuming the typical scale for the radial variation is $R$, the angular momentum equation yields the viscous / accretion timescale:
\begin{equation}
t_{\rm vis} \sim {R^2 \over \nu} \sim \left( 
\frac{T_{\rm g}}{T} \right) t_{\rm dyn}\approx 4{\rm ~days~~} \alpha^{-4/5} \dot{M}_{16}^{3/10} M_1^{1/4} R_{10}^{5/4}
\end{equation}
This is also the characteristic time $R/v_r$ needed to accrete viscously matter initially at a radius $R$ (see \S\ref{evolution}). This accretion timescale is much longer than the dynamical timescale. Perturbations to the hydrostatic balance propagate at the sound speed so the vertical timescale is:
\begin{equation}
t_{\rm vert} \sim {H \over c_{\rm s}} \sim t_{\rm dyn}
\end{equation}
The characteristic time needed to establish hydrostatic balance is of the same order as the dynamical timescale i.e. short compared to the accretion timescale. The energy equation yields the thermal timescale:
\begin{equation}
t_{\rm ther} \sim {e \over q^+} \sim {c_{\rm s}^2  \over {\nu \Omega_{\rm K}^2}} \sim \frac{1}{\alpha} ~t_{\rm dyn}
\end{equation}
This is the characteristic timescale needed to establish the local balance of heating and cooling and is short compared to the accretion timescale.

\subsection{Diffusion equation \label{evolution}}
The mass and angular momentum conservation equations of a thin disc, rewritten in terms of $\Sigma$, can be combined into a single evolution equation:
\begin{equation}
\pd{\Sigma}{t} = \frac{3}{R} \pd{}{R} \left[R^{1/2} \pd{}{R}\left( \nu \Sigma R^{1/2} \right) \right]
\end{equation}
This diffusion equation describes how the column density $\Sigma$ evolves from a given initial radial profile and is discussed in \citet{lp74}. 

The most pedagogical example is that of an infinitely thin annulus of matter which spreads both in and out under the influence of a constant viscosity $\nu$ \citep[Figs.~4-5 of][]{lp74}.  After an initial adjustment, the mass accretion rate onto the compact object approaches a self-similar regime where $\dot{M}$ varies as some power of time $t$. This is characteristic of constant total angular momentum solutions. After a few viscous timescales $t_{\rm vis}\sim R^2/\nu$, most of the initial mass is accreted onto the compact object while a decreasing fraction of the matter carries away the angular momentum at infinity. In a binary, the thin disc has an outer radius set by tidal torques and angular momentum is given back to the companion \citep{pp77}. Semi-analytical solutions to this diffusion equation under different assumptions can also be found in e.g. \citet{bp81} and \citet{ls87}.

\subsection{Stability of steady thin discs \label{stab}}

A steady thin disc is susceptible to thermal and viscous instabilities \citep{ss76,pringle76}. The first arises when cooling cannot keep up with heating i.e. if heating increases faster than cooling when the temperature of an annulus is raised (the thermal balance is local). The stability criterion is:
\begin{equation}
\left(\pd{Q^+}{T}\right)_\Sigma< \left(\pd{Q^-}{T}\right)_\Sigma
\end{equation}
The thermal timescale is much shorter than the viscous timescale on which $\Sigma$ evolves (\S\ref{time}). Setting $\Sigma$ constant, $Q^+\sim \nu\Sigma \sim T$ (Eq.~\ref{qplus}). For optically thick radiative cooling then (Eq.~\ref{thick}) $Q^-\sim T^4 / \tau \sim T^{15/2} H/ \Sigma^2\sim T^8$ (taking Kramers opacity $\tau\sim \rho_o T^{-7/2} \Sigma$), hence the disc of \S\ref{vertstruct} is thermally stable. However, if the medium is optically thin then $Q^-\sim H \rho_o^2 T^{1/2} \sim 1 $ (again for free-free radiation) and the annulus is unstable \citep{prp}. 

The disc is viscously unstable when perturbations of the density are amplified rather than smoothed out by accretion. The criterion for viscous stability is:
\begin{equation}
\pd{\dot{M}}{\Sigma}> 0 {\rm ~~or~~} \pd{\nu \Sigma}{\Sigma}>0
\label{viscousstability}
\end{equation}
using $\dot{M}\sim\nu\Sigma$ (steady-state thin disc). The criterion is easily found by linearizing Eq. 2.11 for small perturbations. Density perturbations grow on the long viscous timescale so thermal equilibrium can be assumed. Since $Q^+\sim \nu \Sigma \Omega_{\rm K}^2 \sim P H \Omega_{\rm K}$ and $Q^- \sim T^4/ \kappa \Sigma\sim P/\Sigma$ one has $H\propto 1/\Sigma$ in a radiation pressure / electron scattering ($\kappa$ constant) dominated annulus. Writing the hydrostatic equilibrium shows $P\sim (\Sigma/H) c_s^2 \sim  \Sigma H \Omega_{\rm K}^2 \propto 1$ is independent of $\Sigma$ hence $\nu \Sigma \propto PH \propto 1/\Sigma$: the annulus is unstable to density perturbations \citep{le74}. This region is also thermally unstable \citep{ss75}.

As mentioned in \S\ref{vertstruct}, the inner region of a steady thin disc at high $\dot{M}$ is radiation pressure and electron scattering dominated. Hence the region from which X-rays originate is unstable and much effort went into finding how to reconciliate this fact with observations. The Shapiro-Lightman-Eardley (SLE) model is one such attempt in which the viscously unstable region gives way to a hot, optically thin plasma at small radii \citep{sle}. In the SLE model, the protons are heated by viscous dissipation and cool by Coulomb interactions with electrons. The electrons lose their energy efficiently by local inverse Compton scattering on soft photons (explaining the observed hard X-rays) so they have a much lower temperature ($T_{\rm e}\sim 10^{8}$~K) than the protons ($T_{\rm p}\sim 10^{11}$~K). This solution is still thermally unstable \citep{pringle76}. Lowering the temperature cools the flow back to the Shakura-Sunyaev solution. Raising the temperature increases viscous heating and decreases the inflow timescale so that protons do not have time to give their energy to the electrons before being accreted. Energy is advected and the SLE flow becomes a two temperature advection-dominated flow (see \S3.2).

The conclusion that a radiation pressure dominated thin disc is unstable depends upon the assumptions made. The region can be stabilised by invoking a different viscosity law \citep{piran}, for instance one in which $\nu$ is proportional to the gas pressure only and not the total pressure ($\beta$ viscosity). Other possibilities include e.g. mass depletion by a wind \citep{piran} or irradiation heating from the boundary layer \citep{czerny}. Most importantly, including radial advection of heat stabilizes the flow at higher accretion rates to give a {\em slim accretion disc} \citep[see \S3.2 below and][]{slim}.

\subsection{The disc instability model \label{dim}}

\begin{figure}
\centerline{\epsfig{file=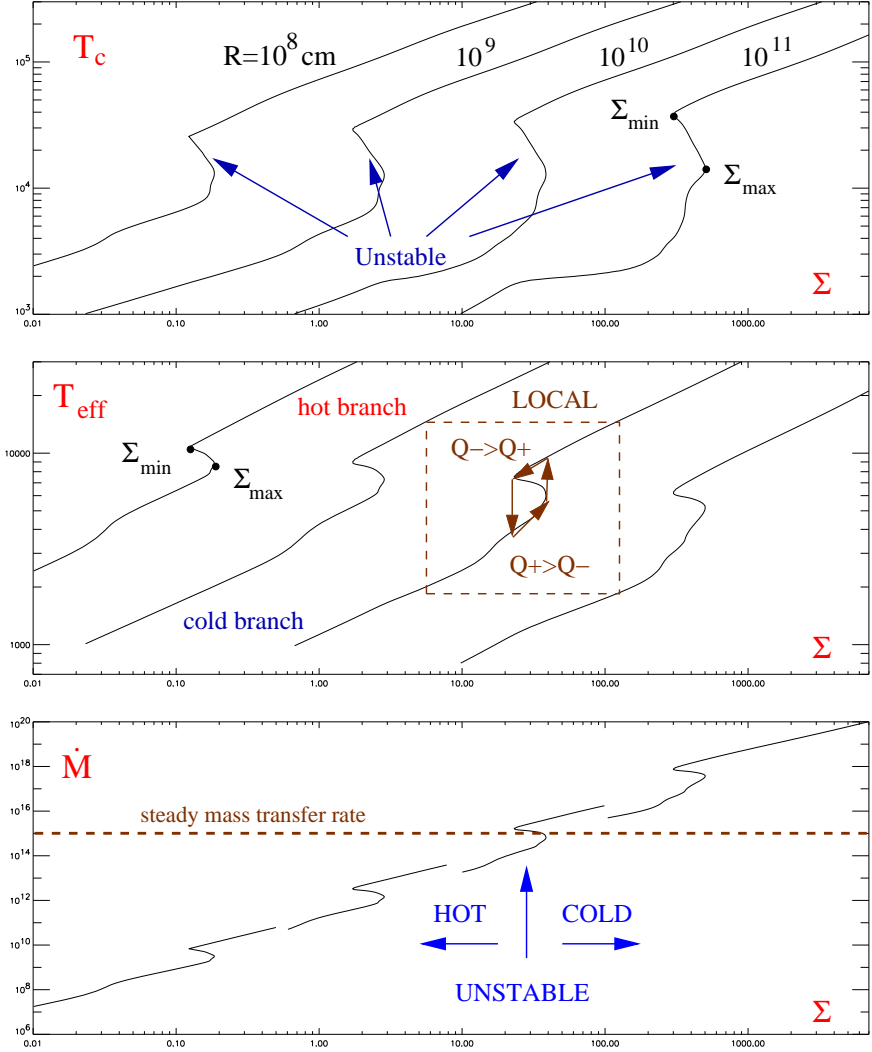,width=12cm}}
\caption{Thermal equilibrium curves $Q^+=Q^-$ calculated at different
radii with $M_\star=7M_\odot$ and $\alpha=0.1$. From top to bottom,
the S curves are represented in the $(\Sigma,T_{\rm c})$,
$(\Sigma,T_{\rm eff})$m and $(\Sigma,\dot{M})$ planes (equivalent in
steady-state). The middle branch is unstable due to hydrogen
recombination and the local limit cycle is shown in the middle
panel. For a large enough disc fuelled by a steady mass transfer rate
the inner region is on the hot branch while the outer is on the cold
branch (easily seen by looking at the intersection between the fixed
$\dot{M}_{rm t}$ and the S curves in the last panel). The intermediate
region is unstable and numerical studies of the non-linear outcome
show the disc cycles between a hot (outbursting) state and a cold
(quiescent) state.}
\end{figure}

The stability of a disc can be studied by looking at the local thermal
equilibrium curves. This is the set of $(\Sigma,\dot{M})$ such that
$Q^+=Q^-$ for given $M_\star$, $\alpha$ and radius $R$ in a
steady-state disc. In steady-state, one can use indifferently $T_{\rm
eff}$ or $T_{\rm c}$ instead of $\dot{M}$ (Eq.~\ref{thick}). If
$\partial{\dot{M}}/\partial{\Sigma}<0$ somewhere along this curve then
the annulus is unstable. This does not happen in the Kramers opacity
regime (\S\ref{stab}). But at temperatures lower than 10$^4$~K the
opacity decreases dramatically due to hydrogen recombination. This
reverses the slope of the $(\Sigma,\dot{M})$ relation until the
temperature has dropped enough that hydrogen is neutral: there is a
range of accretion rates for which the local annulus is unstable
\citep{meyer}. Fig.~1 shows examples of thermal equilibrium curves
calculated at several radii using a detailed vertical integration
\citep[$M_\star=7M_\odot$, $\alpha=0.1$; the code is described
in][]{hameury}.

Such a curve, drawing an S in the $(\Sigma,\dot{M})$ plane, leads to a
limit cycle \citep{bp82}. When mass transfer rates $\dot{M}_{\rm t}$
is in the unstable range, the annulus must either jump up on the top
branch or down on the bottom branch. On the top branch the accretion
rate is higher than the average $\dot{M}_{\rm t}$ so density
decreases. When the density reaches $\Sigma_{\rm min}$, a slight
temperature perturbation causes the annulus to cool on a thermal
timescale to the lower branch ($Q^- > Q^+$ in the region around
$\Sigma_{\rm min}$). The accretion rate is lower than $\dot{M}_{\rm
t}$ on this branch so the density increases until $\Sigma_{\rm max}$
is reached, the annulus heats to the upper branch and the process
repeats itself. Changes in $\Sigma$ occur on the viscous timescale
$t_{\rm vis}\propto 1/T_{\rm c}$ which is shorter on the cold branch
than on the hotter upper branch. The local cycle consists of a rapid
rise to a brief high $\dot{M}$ state followed by a longer low
$\dot{M}$ quiescent state. In a full disc the situation is complicated
since mass is transfered from and to each region. The transitions in the
$(\Sigma,\dot{M})$ plane are then more complex than the above picture
suggests.

The critical values obtained from vertical structure calculations (Fig.~1) are well approximated by fits:
\begin{eqnarray*}
& &\Sigma_{\rm min}\approx 8.3~ \alpha^{-0.7} M_1^{-0.4} R_{10}^{1.1}~{\rm g~cm}^{-2}\\
& &\Sigma_{\rm max}\approx 13.4~ \alpha^{-0.8} M_1^{-0.4} R_{10}^{1.1}~{\rm g~cm}^{-2} \label{sigmamax}\\
& &\dot{M}_{\Sigma_{\rm max}}\approx 4\cdot10^{15}~ \alpha^{0.0} M_1^{-0.9} R_{10}^{2.7}~{\rm g~s}^{-1}\\
& &\dot{M}_{\Sigma_{\rm min}}\approx 9\cdot10^{15}~ \alpha^{0.0} M_1^{-0.9} R_{10}^{2.7}~{\rm g~s}^{-1}
\end{eqnarray*}
For a steady disc to be everywhere on the hot branch (hence stable) requires $\dot{M}_{\rm t}>\dot{M}_{\Sigma_{\rm min}}(R_{\rm disc})$, giving values which are not unrealistic for XRBs and CVs. On the other hand, a stable steady {\em cold} disc requires $\dot{M}_{\rm t}<\dot{M}_{\Sigma_{\rm max}}(R_\star)$ which, due to the steep dependence on $R$, is very low: $\sim$ 10 kg/s for a black hole/neutron star and $\le 10^{13}$~g/s for a white dwarf. Cold, stable steady discs extending down to the surface of the compact object are unlikely to exist.

At intermediate accretion rates, the steady disc solution goes through the unstable branch at some radius. A density front appears and the disc evolves. The front speed is set by the radial inflow rate of matter (Eq.~\ref{angular}):
\begin{equation}
v_r=-\frac{3}{\Sigma R^{1/2}} \pd{}{R} (\nu\Sigma R^{1/2}){\rm ~~so~~~} v_f\approx \frac{\nu}{H} \approx \alpha c_s
\label{vr3}
\end{equation}
since the typical scale of the variations is $\Delta R\sim H$. The propagation timescale is $t_f= R/v_f\sim (t_{\rm vis}t_{\rm ther})^{1/2}$, reflecting that this is a combination of the thermal and viscous instabilities. The disc enters a hot state until it cannot sustain the high accretion rate needed. A cooling front then propagates, shutting off accretion onto the central object and matter piles up until the cycle can start again.

This {\em disc instability model} (DIM) has been applied to CVs, XRBs, AGNs and protostellar discs \citep{hk96,burderi,lasota}. S curves have been calculated by many different groups using various assumptions and degrees of complexity for heat transport or viscous dissipation. Although it should really be confirmed from ab initio calculations of viscous transport, the reversal of the $\nu\Sigma$ slope is a robust unavoidable feature. The location of the critical $\Sigma$ points can vary from model to model: this is equivalent to (small) variations in the unknown parameter $\alpha$. But the critical accretion rates are independent of $\alpha$ as (a) H ionisation is what ultimately sets the critical (surface) $T_{\rm eff}$ so it is approximately constant (around 8000~K for the hot branch, see Fig.~1); and (b) $\dot{M}$ is related to $T_{\rm eff}$ independently of $\nu$ (Eq.~\ref{teff}). Therefore, the essential feature of the DIM is a strong prediction of which systems should be stable. The second important aspect is that it offers a framework to understand the outbursts of unstable systems. In the following, I discuss the major results obtained from the application to cataclysmic variables and X-ray binaries \citep[and defer to][ for a thorough review and complete references]{lasota}.

\subsection{Models of Cataclysmic Variables \label{cv}}

The DIM reproduces well the distinction between novae-like (steady) and dwarf novae (unsteady) systems for cataclysmic variables \citep{smak,osaki}. The DIM is also able to roughly reproduce the outbursts of dwarf novae (confirming the idea of \citealt{osaki2}) but on one condition: that the viscosity on the hot branch is much higher than on the cold branch. Although $\nu$ varies with temperature, a constant $\alpha$ produces only small short outbursts because $\Sigma_{\rm min}$ and $\Sigma_{\rm max}$ are too close. To explain the amplitudes and timescales, $\alpha$ must change with $\alpha_{\rm hot}\sim 0.1$ and $\alpha_{\rm cold}\sim 0.01$ \citep{smak84}. This is one of the most solid conclusions to come out of the DIM which can be tested against theoretical models of angular momentum transport. In the MRI case such a change is not inconceivable as its strength depends upon the gas ionisation fraction. Outbursts can therefore give some constraints on the viscosity, something steady state models cannot do.

However, further investigation shows that many additional effects need to be considered when modelling outbursts. For instance, the DIM assumes a steady mass transfer rate from the secondary but there is no reason for this to be the case. In fact, magnetic CVs do show variations of factors 10-100 which can be directly attributed to $\dot{M}_{\rm t}$ since they have no accretion discs. White dwarfs with lower magnetic moment can also disrupt the inner regions of a quiescent low $\dot{M}$ disc. Tidal torques from the secondary are important in setting the outer disc radius, how it varies and determining the thermal balance in the outer regions. The impact of the stream of matter incoming from the Lagrange point and UV radiation from the hot white dwarf can also significantly heat the disc and change its properties \citep{buat}. A wide variety of CV lightcurves can be accounted for by varying combinations of these effects, using standard values for $\alpha$ and some earlier 'failures' of the DIM were actually due to missing some of these important phenomena \citep{lasota}. 

However, there are still many unresolved issues and this testifies to
large number of high-quality observations of CVs. For instance, the
DIM cannot readily explain the flat quiescent lightcurves or some of
the variations observed in the outbursts from cycle to cycle
\citep{smakox}. Moreover, the DIM still fails to make accurate
spectral predictions, with fits requiring very large values of
$\alpha$. Oversimplification of the radiative transfer may be partly
responsible but the issue really is that the vertical distribution of
the heat generated by angular momentum transport is unknown. This has
a major impact on spectral line formation. A related problem not
addressed by the DIM is the formation of (line-driven) disc winds that
can carry away sizeable fractions of the accreted matter.

\subsection{Models of low mass X-ray binaries}
Low mass X-ray binaries are very similar to CVs, with comparable
companions, orbital periods and Roche lobe overflow mass transfer
rates. One would therefore expect the DIM to work just as well as in
CVs. But the standard DIM wrongly predicts that all observed systems
should be transient. The solution to this problem is that one cannot
neglect irradiation heating in X-ray binaries. High energy radiation
coming from the inner regions of the flow can interact with the flow
at larger radii and heat it to the upper branch of the S curve. A
small fraction ($\sim 0.1$\%) of the luminosity deposited this way is
enough to make a difference. Although there is ample evidence for
irradiation in XRBs \citep{xrbs}, the geometry and processes through
which high energy photons reach the outer disc are still unknown. They
could involve a scattering corona, disc warps and/or a jet (as in the
AGN lamppost model; \citealt{dubusirr}).

A typical dwarf nova has eruptions lasting a few days and recurrence
times of a few months. A typical transient XRB has an eruption lasting
a few months with a recurrence period of years \citep{csl}. Since the
main difference with CVs is not the mass of the accretor but its
radius, one would think that the reason for the different outbursts
has something to do with the flow reaching in to radii $10^{6-7}$~cm
in XRBs instead of 10$^{8-9}$~cm. But numerical simulations show such
discs undergo large numbers of small amplitude, short duration and
recurrence time outbursts. This is not surprising as
Eq.~\ref{sigmamax} shows that the amount of mass needed to trigger an
outburst (i.e. for $\Sigma>\Sigma_{\rm max}$) becomes very small at
small $R$. Including irradiation in outburst lengthens the
outbursts by keeping the disc hot longer but the models are still far
from observed lightcurves. Another problem is that the accretion
rate onto black hole has to be $\lta 10$~kg/s in quiescence since
$\Sigma$ must be lower than $\Sigma_{\rm max}$ everywhere in the
disc. But black hole XRBs are detected at levels of $10^{31}$
erg~s$^{-1}$ which is incompatible with the accretion rate and the
predicted $T_{\rm eff}\approx 3000$~K in quiescent thin discs.

As it turns out, there is an elegant solution to {\em both} of these
problems if the thin disc is assumed to be replaced (`evaporated') by
a low radiative efficiency accretion flow (LRAF) at small radii in
quiescence: (1) this removes the highly sensitive region which
triggered many outbursts; (2) higher accretion rates can be 'hidden'
in the radiatively inefficient flow and help increase the recurrence
timescale (the disc acts as a pierced bucket trying to fill up); (3)
spectral models of a class of such flows can explain the spectral
energy distribution of quiescent XRBs. LRAFs are discussed in the next
section.

Models of a modified DIM including irradiation and the evaporation of
the inner thin disc show good agreement with the observed lightcurves
\citep{dubus}. These models do not require a different viscosity than
that of CVs. This is gratifying since, after all, the conditions in
the disc at large radii are not different in CVs and XRBs. On the
other hand, the prospect of obtaining strong constrains on viscosity
by comparing theoretical and observed lightcurves of CVs and XRBs
fades away as increasingly complex (and poorly known) physics has to
be added to the model.

\section{Low radiative efficiency accretion flows: thick discs}

The obvious contrast to the thin disc approach is to set $f\approx 1$
i.e. to look for solutions with little or no radiation cooling to
Eqs.~\ref{mass}-\ref{hydro}. The low radiative efficiency accretion
flow (LRAF) obtained is cooled by the advection of viscous heat
\citep{ichimaru,katz,b1978,rees,slim,ny94,nyi,acklr}.
Being cooled by advection, LRAFs therefore have a radiative timescale
which is longer than the accretion timescale: $t_{\rm acc}/ t_{\rm
rad} \sim (1-f) (R\Omega /c_s)^2 \ll 1$ when $f=1$. Inversely, this
ratio is very large (as expected) for the thin disc ($f=0$) so there
is a critical parameter value (e.g. $\dot{M}$ at $M_\star$ and
$\alpha$ fixed), set by the detailed radiative processes, beyond which
LRAFs can exist (see \S3.2).

LRAFs are necessarily hot flows with $T\lta T_{\rm g}$ (see
Eqs.~\ref{radialadim}-\ref{eneradim}). A special case is that in which
$f=1$ and $\Omega=0$ which corresponds to a purely radial accretion
flow i.e. to the Bondi-Hoyle-Lyttleton solution describing accretion
from the interstellar medium \citep{bh44}. In this sense, the
adiabatic flows discussed below are a generalisation of this solution
to the case of non-zero initial angular momentum.

Because of their high temperatures and low radiative efficiencies
LRAFs provide an ideal context to model low luminousity objects
showing emission at high energies \citep[e.g.][]{ngc4258} and have
thus attracted considerable interest. Since $T\lta T_{\rm g}$ implies
$H\lta R$, adiabatic flows are geometrically thick and more amenable
to numerical simulations than thin discs in which radiation processes
have to be taken into account and where resolving the very small
vertical scale height is numerically challenging. Theoretical work on
LRAFs has thus been enriched by the possibility to simulate adiabatic
accretion, including ab initio transport of angular momentum via the
MRI (hence requiring no additional assumptions).

The equations for a steady state accretion flow with an $\alpha$ type
viscosity can be put in dimensionless form as shown in \S1.6. Far from
the boundary conditions, the equations have no length scale and a
self-similar solution in which the variables are written in powers of
$R$ is possible. The thin disc is one special case with $f\ll 1$ and
the next section describes the extension to arbitrary $f$
\citep{spruitetal,ny94}. In principle $f$ is found self-consistently
from the radiation transfer and this is done in a rudimentary way in
\S3.2, leading to the ``unified description of accretion flows''. Some
applications are presented in \S3.3 while current problems and
controversies in LRAFs are summarized in \S3.4.

\subsection{The self-similar solution}

Assuming $f$ is constant, Eqs.~\ref{radialadim}-\ref{eneradim} immediately show that $\Omega\propto \Omega_{\rm K}\propto R^{-3/2}$ and $T\propto T_{\rm g}\propto R^{-1}$. Using these in Eq.~\ref{vr} gives $v_r\propto R^{-1/2}$. The hydrostatic balance $H\Omega_{\rm K}=c_s\propto T^{1/2}$ gives $H\propto R$, mass conservation gives $\Sigma\propto R^{-1/2}$ so $\rho_o\propto R^{-3/2}$. With these dependences, Eq.~\ref{radialadim}-\ref{eneradim} solve for $\Omega$ and $T$ from which the other variables can be deduced:
\begin{eqnarray}
\Omega^2 &=&(\epsilon/f) ~h(\alpha,\gamma,f)~\Omega^2_{\rm K}\\
v_r &=& -(3/2)\alpha ~h(\alpha,\gamma,f)~R\Omega_{\rm K}\\
T &=& (3/2)(1+\epsilon)~h(\alpha,\gamma,f)~T_{\rm g}\\
H/R &=& f/\epsilon
\end{eqnarray}
where $\epsilon=(5/3-\gamma)/(\gamma-1)$ and $h$ is a function\footnote{$h(\alpha,\gamma,f)$ is the solution to $9\alpha^2 h^2 + 4(5+2\frac{\epsilon}{f}) h -8=0$ } of $\alpha$, $\gamma$ and $f$. To fix ideas, $\Omega\approx 0.7 \Omega_{\rm K}$, $v_r\approx 5\cdot 10^7$ cm$\cdot$s$^{-1}$, $T\approx 4\cdot 10^9$~K and $H/R\approx 0.75$ at $10^9$~cm around a 10~M$_\odot$ black hole with $f=0.75$, $\alpha=0.1$ and $\gamma=4/3$. The flow is sub-Keplerian, very hot and the accretion timescale is very short compared to the thin disc.

The thin disc is recovered as expected when $f\rightarrow 0$, giving $h\approx f/\epsilon \rightarrow 0$.  In the other limit, as $f$ increases towards 1, $h$ increases so the temperature, viscosity and inflow velocity $v_r \sim \nu/R$ all increase. Since $t_{\rm acc}=R/v_r\sim (1/h) t_{\rm ther}$, the radial inflow time can become comparable to the thermal time. The disc becomes sub-Keplerian and thick with $H/R\lta 1$. If the gas is adiabatic with $\gamma=5/3$, the flow becomes purely radial with $\Omega\rightarrow 0$ i.e. the solution tends to Bondi spherical accretion. \cite{ogilvie} studied the evolution of an initial torus of gas with some angular momentum at $R$ accreting adiabatically, a setup similar to that described for thin discs in \S2.5. He finds the self-similar solution with $f=1$ is asymptotically approached with time.

There are some obvious limitations to the self-similar solutions. The self-similar solution extends at all $R$ and does not address what happens at boundaries. But integration of the equations with physically motivated boundary conditions show the self-similar solution describes the flow well enough far from the boundaries \citep{cal,nkh}. Global solutions to the equations in general relativity have also been found \citep{acgl,pa97,pg98}. The validity of using the vertically integrated set of equations is not obvious when $H/R\sim 1$. \cite{ny2d} studied self-similar solutions to the equivalent 2D ($R$,$\theta$) set of equations and found that the above solution was a good approximation if the integration is interpreted in terms of $\theta$ instead of $z$.

\subsection{Radiative processes: ADAFs and slim discs}
The parameter $f$ was assumed given when it should really be found self-consistently from the radiative processes at all radii. In practice, this implies guessing $f$, determining $Q^-$ for all relevant radiative processes (e.g. optically thick flux or free free, synchrotron, pair creation, comptonisation etc.) using the values for $\Sigma$ and $T$ from the radial solution, work out the improved value for $f$ and repeat the process until convergence is reached \citep[e.g.][]{nyi}.  To complicate things further there is no a priori reason for $f$ to be constant in the flow but this is an assumption which is often made for simplicity and is likely to be reasonable in light of other approximations in the radiative transfer \citep{esin}.

\subsubsection{Optically thin case: ADAFs}
Here, all of these complications are left aside and simple rudimentary forms for $Q^-$ are used to find self-consistent solutions. The energy equation (Eq.~\ref{energy}) in steady state can be rewritten as \citep{acklr}:
\begin{equation}
Q_{\rm adv}=Q_{\rm vis}^+-Q^- \rightarrow \xi \left(\frac{\dot{M}}{2\pi}\right)^2 \frac{\Omega_{\rm K}}{\alpha \Sigma R^2} =\Omega^2 \omega^2 \left(\frac{\dot{M}}{2\pi}\right)+\omega Q^-
\label{mdot2}
\end{equation}
where $\omega={\rm d}\ln \Omega/{\rm d}\ln R$ and $\xi={\rm d}\ln \rho T^{1/1-\gamma}/{\rm d}\ln R+4(1-\beta){\rm d}\ln \rho T^{-3}/{\rm d}\ln R$ for a mixture of perfect gas and radiation ($\beta$ is the ratio of gas pressure to total pressure). In an adiabatic flow $Q^-=0$ so the equation reduces to $\xi=\omega (R\Omega/c_s)^2$. Adiabatic accretion is possible for any $\dot{M}$ when this is verified.

If $Q^-$ is not negligible and the flow is emitting optically thin free-free radiation then $Q^-\sim H\rho_o^2 T^{1/2}\sim \Omega_{\rm K}\Sigma^2$ ($\beta=1$). For a given $R$, $\Sigma$ and $\alpha$ the mass accretion rate $\dot{M}$ satisfying the thermal equation (Eq.~\ref{mdot2}) depends on global properties of the flow. The thermal equation can be made local by assuming that the self-similar scalings hold and that $\Omega=\Omega_{\rm K}$. Eq.~\ref{mdot2} then becomes $\dot{M}^2-\alpha C_1 \Sigma R^{1/2} \dot{M} + \alpha C_2 \Sigma^3 R^2=0$ ($C_1$, $C_2$ constants). Below a critical $\Sigma_{\rm crit}\propto \alpha/R$ there are two possible solutions for $\dot{M}$: one with $Q^->Q_{\rm adv}$ corresponding to a Shapiro-Lightman-Eardley flow (see \S\ref{stab}) and the other with $Q_{\rm adv}>Q^-$ is called the advection-dominated accretion flow (ADAF) in the literature. Above the critical $\dot{M}_{\rm crit}\propto \alpha^2 R^{-1/2}$ corresponding to $\Sigma_{\rm crit}$ there is no solution. $\dot{M}_{\rm crit}$ can also be found by equating the accretion timescale to the radiative timescale for bremstrahlung: ADAFs appear when the latter becomes longer than the former. 

\subsubsection{Optically thick case: slim discs}
In the optically thick case we have already seen that one solution is the Shakura-Sunyaev thin disc solution ($Q^+=Q^-$). As $\dot{M}$ increases, radiation pressure increases and the solution becomes unstable (\S\ref{stab}). Writing the cooling term in the Eddington approximation with electron scattering opacity and assuming radiation pressure dominates, the thermal equilibrium can be rewritten with $H/R$ instead of $\dot{M}$ as:
\begin{equation}
Q_{\rm adv}=Q_{\rm vis}^+-Q^- \rightarrow  \xi \left(\frac{H}{R}\right)^3 = \left(\frac{H}{R} \right) \frac{\Omega^2}{\Omega_{\rm K}^2} -\frac{C}{\alpha \omega R \Omega_{\rm K} \Sigma}
\label{sc2}
\end{equation}
where $C$ is a constant. Again, this equation can be made local by assuming Keplerian rotation and $\xi$ constant. For a given $\Sigma$, $R$ and $\alpha$ there can be up to three solutions for $H/R$. Detailed calculations show an S-curve in the ($\Sigma$,$H/R$) plane where the bottom branch is the Shakura-Sunyaev solution, the middle branch the unstable $P_{\rm rad}$ dominated thin disc and the top branch a new solution called the {\em slim disc} solution. Slim discs are radiation-pressure dominated and cooled by advection. 

\subsubsection{Thermal equilibrium curves}
Both the ADAF and slim disc branches are thermally and viscously
stable \citep[see][ and references therein for
details]{honma,kato}. An easy way to see this is to plot the different
branches found above in the ($\Sigma$,$\dot{M}$) in a similar fashion
to what was done for the S-curves of the disc instability model in
\S\ref{dim}. This is shown in Fig.~2 (adapted from
\citealt{chen95}). At low $\Sigma$, the disc is optically thin and
there are two solutions: the thermally unstable SLE solution and the
stable ADAF branch. At higher $\Sigma$, the disc is optically thick
and there are three branches: the Shakura-Sunyaev thin
disc (i.e. the upper branch of the S-curve in Fig.~1; the S
corresponding to hydrogen ionisation occurs at much lower values of
$\dot{M}$ than shown in Fig.~2; see \S2.7.), the radiation pressure
dominated thermally/viscously unstable thin disc and the slim disc
branch at high $\dot{M}$. There is a change of topology at a critical
$\alpha$ above which an advection-dominated solution is possible for
all $\dot{M}$. Although these curves are a useful tool to illustrate
the various accretion solutions, it is worth stressing that they are
local representations of an inherently non-local phenomenon
(advection) and that they assume an $\alpha$ type viscosity. These
curves therefore are not in any way a complete description of
accretion flows.
\begin{figure}
\centerline{\epsfig{file=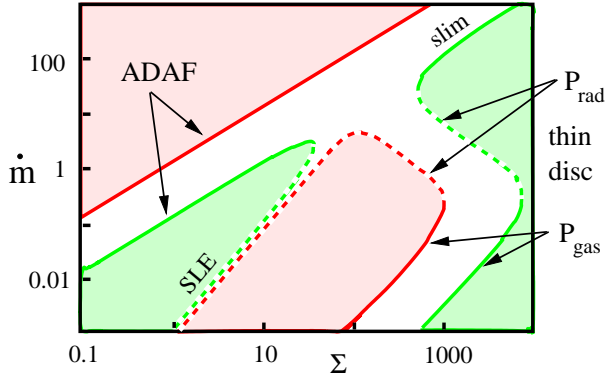,width=8cm}}
\caption{A global view of the thermal equilibrium curves in the column density $\Sigma$, mass accretion rate $\dot{m}$ (normalised to the Eddington rate) plane. This figure, adapted from \citet{chen95}, shows the branches corresponding to different types of accretion. The curves are calculated at a radius of 30 $R_g$ for a black hole of 10~M$_\odot$ and using $\alpha=0.1$ (green/light lines) or $\alpha=1$ (red/dark lines). The dark areas enclose the regions where $Q^+>Q^-$ in the case $\alpha=0.1$ and $Q^->Q^+$ in the case $\alpha=1$. This shows the SLE solution (dotted lines) is thermally unstable in both cases (a slight increase in temperature at constant $\Sigma$ pushes a disc on the SLE branch in the $Q^+>Q^-$ region so that it cannot cool). On the other hand, the ADAF, slim disc and gas pressure dominated thin disc branches are stable. The radiation pressure dominated thin disc (dotted line) is thermally and viscously unstable.}
\end{figure}

\subsection{Applications of LRAFs}

LRAFs have applications to a variety of different objects in various situations. The S-curve at the transition between a thin disc and a slim disc (see Fig.~2), when radiation pressure takes over, opens up the possibility of a thermal-viscous cycle like that described in \S\ref{dim}. The associated mass accretion rates are very high and the timescales are very short so the cycle timescale is fast, of the order of seconds to minutes. This has found possible applications in some XRBs which accrete at high rates \citep[e.g.][ and references therein]{cannizzoslim,szu}.

Optically thin LRAFs have many more applications. As mentioned at the end of \S2, models of transient XRBs suggest a LRAF replaces the thin disc at small radii in quiescence. Support for this comes from models of optically thin LRAFs that reproduce the spectral energy distribution of quiescent X-ray binaries \citep{nmyi96}. A comparison of the quiescent luminosities of neutron star transient XRBs against black hole transient XRBs shows that the neutron star systems are systematically more luminous by an order of magnitude or two than their black hole counterparts. This has been interpreted as evidence for a horizon in black holes. In both cases the compact object is surrounded by a LRAF. However, the energy stored in the flow is radiated when it reaches the surface of the neutron star while it disappears through the horizon in a black hole, explaining the luminosity difference \citep[for a recent discussion see][]{garcia}.

The \cite{esin} model of transient XRBs spectral states is a natural extension of the idea that LRAFs and thin discs coexist. As the transient XRB goes into outburst, the ADAF retreats to smaller radii and eventually is fully replaced by the thin disc when the increasing mass flow rate is above the ADAF critical mass accretion rate (\S3.2). Since the thin disc has a much lower temperature than the ADAF this roughly implies a transition between a hard and a soft X-ray spectrum during the rise to outburst and vice-versa during the return to quiescence. These ideas have been applied to the observed spectral changes in Nova Mus 1991 \citep{esin}.

However, the most studied application of LRAFs is to Sgr A*, the black hole at the center of our Galaxy. Despite its dynamically-measured mass of a few 10$^6$~M$_\odot$, the X-ray luminosity of Sgr A* is only $10^{33}$ ergs~s$^{-1}$, well below the Eddington luminosity. The mass accretion rate inferred from observations is much higher than the luminosity suggests making it an ideal candidate for a LRAF. The dormant massive black holes at the center of nearby galaxies are also prime candidates for LRAFs \citep[for a review of the above see][]{sgra}.

\subsection{Current issues}

\subsubsection{LRAFs around neutron stars and white dwarfs}
Matter accreting onto a black hole passes necessarily through the last
marginally stable circular orbit and a sonic point before free-falling
through the horizon. The inner boundary condition of global LRAF
solutions around black holes is therefore provided by the regularity
requirement at the sonic point. This sets the specific angular
momentum accreted by the black hole. In contrast, matter falling onto
a neutron star will have a vanishing radial velocity and may not
necessarily pass through a sonic point (this depends on the neutron
star radius i.e. equation of state). The subsonic flow can spin-up or
spin-down the neutron star. Energy released at the surface of the
neutron star can significantly change the luminosity, spectra and heat
budget of the accretion flow. All of these features make accretion
onto a neutron star (or white dwarf) a difficult problem to tackle
\citep[see][ and references therein]{mn01}.

\subsubsection{Two temperature plasmas}
Optically thin LRAF solutions exist only below the critical mass
accretion rate at which the accretion timescale is shorter than the
radiation timescale (\S3.2). Since it is mostly the electrons which
radiate, this mass accretion rate is set by the electron
temperature. To be astrophysically interesting, the electrons must be
assumed to have a much lower temperature than the protons
\citep{sle,rees}. This is achieved if viscous heating mostly goes to
the protons and if these exchange heat with electrons only via Coulomb
collisions (another issue is whether protons and electrons have enough
time to thermalize before being accreted, see \citealt{mahqua}). All
applications have assumed such two temperature LRAFs (the self-similar
solution of \S3.1 assumes a single temperature for simplicity). The
relative fraction $\delta$ of heating that ultimately goes to the
electrons depends on the microphysics and is uncertain
\citep{nyi,beta0,beta1}. Plasma instabilities could erase any
temperature difference between electrons and protons
\citep{beta4,beta3}. Observational constraints can in principle be
obtained since the radiation spectrum depends on $\delta$.

\subsubsection{ADIOS}
Another interesting feature appears when looking at the total local energy of the gas (per unit mass), defined as the sum of the kinetic, internal and potential energy from gravitation and pressure:
\begin{equation}
E=e_c+e_i+e_g+e_p=\frac{1}{2}(v_r^2+R^2 \Omega^2)+e-\frac{GM}{R}+\frac{P}{\rho}=\frac{3\epsilon (f-1/3)}{5f+2\epsilon} R^2 \Omega_{\rm K}^2
\end{equation}
This is also the Bernouilli integral, which is constant along
streamlines for an adiabatic inviscid fluid ($\alpha$=0 and
$f=1$). Since the number is positive for $f>1/3$, a particle going
outwards on a streamline could reach infinity with a net positive
energy, i.e. the fluid is energetically unbound and there could be an
outflow. The local energy is positive because potential energy
liberated at small radii is radiated at larger radii just like in the
thin disc. When the flow cools efficiently this extra energy is
efficiently radiated locally and the Bernouilli number is negative
(which explains the $f=1/3$ limit). Note that in the self-similar
solution there is an infinite amount of energy released at $R=0$ to
redistribute hence it is unclear whether realistic boundary conditions
will also yield a positive Bernouilli integral \citep{ali} .

The assumption that radiatively inefficient flows will necessarily
lead to outflows is referred to as the Advection Dominated
Inflow-Outflow Solution (ADIOS) after \cite{bb}. Work on ADIOS has
focused on extensions of the ADAF with an outflow parameterised by
$\dot{M}\propto R^p$ ($p$ constant). Whether Nature prefers ADIOS to
ADAFs is an open issue with no clear observational tests \citep{ADIOSspec}
and conflicting numerical simulations \citep{hbs2001,in2002}. 

\subsubsection{CDAF}
It may turn out that
neither ADAFs or ADIOS are correct descriptions. ADAFs are unstable
according to the Hoiland criterion for convective
stability of a rotating flow:
\begin{equation}
{\rm ~~stable~if~~}(1-\gamma)\frac{T}{T_g} \Omega_k^2 \left( \fd{\ln P}{\ln R}\right) \fd{\ln (P^{1/\gamma}/\rho)}{\ln R}+\Omega^2\fd{\ln(R^4\Omega^2)}{\ln R} >0
\end{equation}
where the first term is the (squared) Brunt-Vaiasala frequency (the
frequency at which a perturbed blob of gas oscillates around its
position in a stratified atmosphere) and the second term is the
(squared) epicyclic frequency (the frequency at which a perturbed
particle oscillates around its orbit). Convection will transport
angular momentum and energy, changing the radial structure of the
flow. The Convection Dominated Accretion Flow (CDAF) is one solution
which assumes convection will transport angular momentum inwards (by
analogy with axisymetric hydrodynamic flows) and that the steady state
flow is close to the marginally stable limit at which convective
angular momentum transport balances outwards viscous transport. One
condition for this to happen is that viscous transport is not too
strong, i.e. for low values of $\alpha$. There is then a net flux of
energy outwards but very little net accretion (as opposed to ADAFs),
yielding the CDAF self-similar scaling $\rho\propto R^{-1/2}$
\citep{cdaf, cdaf2}. CDAFs have attracted a lot of attention because
this density scaling was found in several numerical simulations
\citep[e.g.][]{spb1999}. CDAFs are highly debated for several reasons:
(1) purely hydrodynamical simulations with a Navier-Stokes viscosity
may not catch the basic behaviour of a realistic flow; (2) MHD
simulations may be inappropriate because the plasma is collisionless;
(3) the Hoiland criterion may be inappropriate; (4) CDAFs may be
thermodynamically impossible since energy dissipated in the MRI
turbulent cascade cannot be used to generate convection \citep[see
e.g.][]{balbuscdaf2,bh2002,in2002,nqia}.

\section{The transition between thin and thick discs}

As illustrated by Fig.~2, accretion with a given $\dot{M}$ has a
choice between several different types of flows. There is at present
no guidance as to which particular type of flow is preferred, which
admittedly an important open question when considering the potential
applications of LRAFs. The flow may always prefer a LRAF whenever it
is possible and this is called the strong LRAF (or ADAF) principle; or
it may choose the LRAF only when there is no other solution available
(the weak principle). One application in which this has important
consequences is the LRAF+thin disc model of transient XRBs. Insights
into this issue might be gained by studying the directly related
problem of the physics of the transition from a thin disc to a LRAF
which is briefly discussed here.

\subsection{Condition for the transition}
Despite its shortcomings, an examination of Fig.~2 suffices to show
most of the problems linked with the transition from a LRAF to a thin
disc. A first possibility is to have a slim disc - thin disc
transition but this occurs at high $\dot{M}$ and there will be an
unstable region (see \S3.3). When $\alpha> \alpha_{\rm crit}$ the thin
disc makes a transition to the Shapiro-Lightman-Eardley type flow
rather than the slim disc. However, the SLE flow is unstable
(\S\ref{stab}). These are the only possible smooth transitions with a
thin disc and both will are unstable. The transition from a thin disc
to an optically thin LRAF (ADAF), which is observationally motivated,
necessarily involves some discontinuity in the framework of Fig.~2. At
the transition, flow variables such as the radial inflow speed $v_r$
or the temperature $T$ must increase by several orders of magnitude
and this involves some additional input of energy which is not taken
into account in Fig.~2.

\subsection{Possibilities}
Studies of the transition between an optically thin LRAF and a thin
disc therefore always involve some additional energy source. In the
vertically averaged model of \citet{honma2} the extra term comes from
turbulent energy transport. This term dominates in the transition
region. The transition is likely to be a 2D or 3D process that occurs
gradually in radius i.e. the thin disc evaporates into a more tenuous
gas in a sandwich around it. The study is therefore not much different
from the studies of disc corona models. The generic idea is that
radiative cooling is $\propto \rho^2$ so that any heating of the
tenuous upper layers of a thin disc can lead to runaway heating (see
\S\ref{stab}) and from there to conditions reminiscent of LRAFs. The
energy terms that can be involved include viscous heating (usually
taken to be $\propto P$ hence $\propto \rho$, \S\ref{stab}), heating
by the reconnection of magnetic loops \citep{galeev}, irradiation by
photons or ions from the inner flow \citep{deufel}, or electron
conduction \citep[][ and references therein]{mlm}. Progress requires a
better understanding of how the energy is dissipated, transported and
radiated, much of which is based on difficult microphysics. This issue
has started to be tackled by MHD numerical simulations \citep{miller}.

\bibliographystyle{aa}
\bibliography{ttdisk}

\begin{thebibliography}{104}
\expandafter\ifx\csname natexlab\endcsname\relax\def\natexlab#1{#1}\fi

\bibitem[{{Abramowicz} {et~al.}(1995){Abramowicz}, {Chen}, {Kato}, {Lasota}, \&
  {Regev}}]{acklr}
{Abramowicz}, M.~A., {Chen}, X., {Kato}, S., {Lasota}, J., \& {Regev}, O. 1995,
  \apjl, 438, L37

\bibitem[{{Abramowicz} {et~al.}(1996){Abramowicz}, {Chen}, {Granath}, \&
  {Lasota}}]{acgl}
{Abramowicz}, M.~A., {Chen}, X.-M., {Granath}, M., \& {Lasota}, J.-P. 1996,
  \apj, 471, 762

\bibitem[{{Abramowicz} {et~al.}(1988){Abramowicz}, {Czerny}, {Lasota}, \&
  {Szuszkiewicz}}]{slim}
{Abramowicz}, M.~A., {Czerny}, B., {Lasota}, J.~P., \& {Szuszkiewicz}, E. 1988,
  \apj, 332, 646

\bibitem[{{Abramowicz} {et~al.}(2000){Abramowicz}, {Lasota}, \&
  {Igumenshchev}}]{ali}
{Abramowicz}, M.~A., {Lasota}, J., \& {Igumenshchev}, I.~V. 2000, \mnras, 314,
  775

\bibitem[{{Balbus} \& {Hawley}(1998)}]{bh}
{Balbus}, S.~A. \& {Hawley}, J.~F. 1998, {Rev. Mod. Phys.}, 70, 1

\bibitem[{{Balbus} \& {Hawley}(2002)}]{balbuscdaf2}
---. 2002, \apj, 573, in press

\bibitem[{{Balbus} \& {Papaloizou}(1999)}]{bp}
{Balbus}, S.~A. \& {Papaloizou}, J.~C.~B. 1999, \apj, 521, 650

\bibitem[{{Baptista}(2001)}]{baptista}
{Baptista}, R. 2001, in Astrotomography, Indirect Imaging Methods in
  Observational Astronomy, Ed. H.M.J. Boffin, D. Steeghs, J. Cuypers, Lecture
  Notes in Physics, vol. 573, 307

\bibitem[{{Bath} \& {Pringle}(1981)}]{bp81}
{Bath}, G.~T. \& {Pringle}, J.~E. 1981, \mnras, 194, 967

\bibitem[{{Bath} \& {Pringle}(1982)}]{bp82}
---. 1982, \mnras, 199, 267

\bibitem[{{Begelman}(1978)}]{b1978}
{Begelman}, M.~C. 1978, \mnras, 184, 53

\bibitem[{{Begelman}(2001)}]{begelman}
---. 2001, \apj, 551, 897

\bibitem[{{Begelman} \& {Chiueh}(1988)}]{beta4}
{Begelman}, M.~C. \& {Chiueh}, T. 1988, \apj, 332, 872

\bibitem[{{Bisnovatyi-Kogan} \& {Lovelace}(1997)}]{beta3}
{Bisnovatyi-Kogan}, G.~S. \& {Lovelace}, R.~V.~E. 1997, \apjl, 486, L43

\bibitem[{{Blandford} \& {Begelman}(1999)}]{bb}
{Blandford}, R.~D. \& {Begelman}, M.~C. 1999, \mnras, 303, L1

\bibitem[{{Blandford} \& {Payne}(1982)}]{blapay}
{Blandford}, R.~D. \& {Payne}, D.~G. 1982, \mnras, 199, 883

\bibitem[{{Bondi} \& {Hoyle}(1944)}]{bh44}
{Bondi}, H. \& {Hoyle}, F. 1944, \mnras, 104, 273

\bibitem[{{Buat-M{\' e}nard} {et~al.}(2001){Buat-M{\' e}nard}, {Hameury}, \&
  {Lasota}}]{buat}
{Buat-M{\' e}nard}, V., {Hameury}, J.-M., \& {Lasota}, J.-P. 2001, \aap, 366,
  612

\bibitem[{{Burderi} {et~al.}(1998){Burderi}, {King}, \&
  {Szuszkiewicz}}]{burderi}
{Burderi}, L., {King}, A.~R., \& {Szuszkiewicz}, E. 1998, \apj, 509, 85

\bibitem[{{Cannizzo}(1996)}]{cannizzoslim}
{Cannizzo}, J.~K. 1996, \apjl, 466, L31

\bibitem[{{Chen} {et~al.}(1997{\natexlab{a}}){Chen}, {Shrader}, \&
  {Livio}}]{csl}
{Chen}, W., {Shrader}, C.~R., \& {Livio}, M. 1997{\natexlab{a}}, \apj, 491, 312

\bibitem[{{Chen} {et~al.}(1997{\natexlab{b}}){Chen}, {Abramowicz}, \&
  {Lasota}}]{cal}
{Chen}, X., {Abramowicz}, M.~A., \& {Lasota}, J. 1997{\natexlab{b}}, \apj, 476,
  61

\bibitem[{{Chen} {et~al.}(1995){Chen}, {Abramowicz}, {Lasota}, {Narayan}, \&
  {Yi}}]{chen95}
{Chen}, X., {Abramowicz}, M.~A., {Lasota}, J., {Narayan}, R., \& {Yi}, I. 1995,
  \apjl, 443, L61

\bibitem[{{Czerny} {et~al.}(1986){Czerny}, {Czerny}, \& {Grindlay}}]{czerny}
{Czerny}, B., {Czerny}, M., \& {Grindlay}, J.~E. 1986, \apj, 311, 241

\bibitem[{{Deufel} \& {Spruit}(2000)}]{deufel}
{Deufel}, B. \& {Spruit}, H.~C. 2000, \aap, 362, 1

\bibitem[{{Dubus} {et~al.}(2001){Dubus}, {Hameury}, \& {Lasota}}]{dubus}
{Dubus}, G., {Hameury}, J.-M., \& {Lasota}, J.-P. 2001, \aap, 373, 251

\bibitem[{{Dubus} {et~al.}(1999){Dubus}, {Lasota}, {Hameury}, \&
  {Charles}}]{dubusirr}
{Dubus}, G., {Lasota}, J., {Hameury}, J., \& {Charles}, P. 1999, \mnras, 303,
  139

\bibitem[{{Esin} {et~al.}(1997){Esin}, {McClintock}, \& {Narayan}}]{esin}
{Esin}, A.~A., {McClintock}, J.~E., \& {Narayan}, R. 1997, \apj, 489, 865

\bibitem[{{Frank} {et~al.}(2002){Frank}, {King}, \& {Raine}}]{fkr}
{Frank}, J., {King}, A., \& {Raine}, D. 2002, {Accretion Power in Astrophysics,
  3rd ed.} (Cambridge University Press)

\bibitem[{{Galeev} {et~al.}(1979){Galeev}, {Rosner}, \& {Vaiana}}]{galeev}
{Galeev}, A.~A., {Rosner}, R., \& {Vaiana}, G.~S. 1979, \apj, 229, 318

\bibitem[{{Gammie}(1999)}]{gammie}
{Gammie}, C.~F. 1999, \apjl, 522, L57

\bibitem[{{Gammie} \& {Menou}(1998)}]{menou}
{Gammie}, C.~F. \& {Menou}, K. 1998, \apjl, 492, L75

\bibitem[{{Garcia} {et~al.}(2001){Garcia}, {McClintock}, {Narayan}, {Callanan},
  {Barret}, \& {Murray}}]{garcia}
{Garcia}, M.~R., {McClintock}, J.~E., {Narayan}, R., {et~al.} 2001, \apjl, 553,
  L47

\bibitem[{{Hameury} {et~al.}(1998){Hameury}, {Menou}, {Dubus}, {Lasota}, \&
  {Hure}}]{hameury}
{Hameury}, J., {Menou}, K., {Dubus}, G., {Lasota}, J., \& {Hure}, J. 1998,
  \mnras, 298, 1048

\bibitem[{{Hartmann} \& {Kenyon}(1996)}]{hk96}
{Hartmann}, L. \& {Kenyon}, S.~J. 1996, \araa, 34, 207

\bibitem[{{Hawley} \& {Balbus}(2002)}]{bh2002}
{Hawley}, J.~F. \& {Balbus}, S.~A. 2002, \apj\ in press (astro-ph/02033309)

\bibitem[{{Hawley} {et~al.}(2001){Hawley}, {Balbus}, \& {Stone}}]{hbs2001}
{Hawley}, J.~F., {Balbus}, S.~A., \& {Stone}, J.~M. 2001, \apjl, 554, L49

\bibitem[{{Honma}(1996)}]{honma2}
{Honma}, F. 1996, \pasj, 48, 77

\bibitem[{{Honma} {et~al.}(1991){Honma}, {Kato}, {Matsumoto}, \&
  {Abramowicz}}]{honma}
{Honma}, F., {Kato}, S., {Matsumoto}, R., \& {Abramowicz}, M.~A. 1991, \pasj,
  43, 261

\bibitem[{{Hubeny}(1990)}]{hubeny}
{Hubeny}, I. 1990, \apj, 351, 632

\bibitem[{{Ichimaru}(1977)}]{ichimaru}
{Ichimaru}, S. 1977, \apj, 214, 840

\bibitem[{{Igumenshchev} \& {Narayan}(2002)}]{in2002}
{Igumenshchev}, I.~V. \& {Narayan}, R. 2002, \apj, 566, 137

\bibitem[{{Kato} {et~al.}(1997){Kato}, {Yamasaki}, {Abramowicz}, \&
  {Chen}}]{kato}
{Kato}, S., {Yamasaki}, T., {Abramowicz}, M.~A., \& {Chen}, X. 1997, \pasj, 49,
  221

\bibitem[{{Katz}(1977)}]{katz}
{Katz}, J.~I. 1977, \apj, 215, 265

\bibitem[{{Lasota}(2001)}]{lasota}
{Lasota}, J.-P. 2001, New Astronomy Review, 45, 449

\bibitem[{{Lasota} {et~al.}(1996){Lasota}, {Abramowicz}, {Chen}, {Krolik},
  {Narayan}, \& {Yi}}]{ngc4258}
{Lasota}, J.-P., {Abramowicz}, M.~A., {Chen}, X., {et~al.} 1996, \apj, 462, 142

\bibitem[{{Lewin} {et~al.}(1995){Lewin}, {van Paradijs}, \& {van den
  Heuvel}}]{xrbs}
{Lewin}, W.~H.~G., {van Paradijs}, J., \& {van den Heuvel}, E.~P.~J. 1995,
  {X-ray binaries} (Cambridge University Press)

\bibitem[{{Lightman} \& {Eardley}(1974)}]{le74}
{Lightman}, A.~P. \& {Eardley}, D.~M. 1974, \apjl, 187, L1

\bibitem[{{Lightman} \& {Shapiro}(1975)}]{ls1975}
{Lightman}, A.~P. \& {Shapiro}, S.~L. 1975, \apjl, 198, L73

\bibitem[{{Loeb} \& {Laor}(1992)}]{loeb}
{Loeb}, A. \& {Laor}, A. 1992, \apj, 384, 115

\bibitem[{{Lubow} \& {Shu}(1975)}]{lubow}
{Lubow}, S.~H. \& {Shu}, F.~H. 1975, \apj, 198, 383

\bibitem[{{Ludwig} \& {Meyer}(1998)}]{ludwig}
{Ludwig}, K. \& {Meyer}, F. 1998, \aap, 329, 559

\bibitem[{{Lynden-Bell} \& {Pringle}(1974)}]{lp74}
{Lynden-Bell}, D. \& {Pringle}, J.~E. 1974, \mnras, 168, 603

\bibitem[{{Lyubarskii} \& {Shakura}(1987)}]{ls87}
{Lyubarskii}, Y.~E. \& {Shakura}, N.~I. 1987, Soviet Astronomy Letters, 13, 386

\bibitem[{{Mahadevan} \& {Quataert}(1997)}]{mahqua}
{Mahadevan}, R. \& {Quataert}, E. 1997, \apj, 490, 605

\bibitem[{{Markoff} {et~al.}(2001){Markoff}, {Falcke}, \& {Fender}}]{mf}
{Markoff}, S., {Falcke}, H., \& {Fender}, R. 2001, \aap, 372, L25

\bibitem[{{Medvedev} \& {Narayan}(2001)}]{mn01}
{Medvedev}, M.~V. \& {Narayan}, R. 2001, \apj, 554, 1255

\bibitem[{{Meyer} {et~al.}(2000){Meyer}, {Liu}, \& {Meyer-Hofmeister}}]{mlm}
{Meyer}, F., {Liu}, B.~F., \& {Meyer-Hofmeister}, E. 2000, \aap, 361, 175

\bibitem[{{Meyer} \& {Meyer-Hofmeister}(1981)}]{meyer}
{Meyer}, F. \& {Meyer-Hofmeister}, E. 1981, \aap, 104, L10

\bibitem[{{Miller} \& {Stone}(2000)}]{miller}
{Miller}, K.~A. \& {Stone}, J.~M. 2000, \apj, 534, 398

\bibitem[{{Narayan}(2002)}]{sgra}
{Narayan}, R. 2002, in Lighthouses of the Universe, ed. {M. Gilfanov, R.
  Sunyaev et al.} (Springer-Verlag), {(astro--ph/02011260)}

\bibitem[{{Narayan} {et~al.}(2000){Narayan}, {Igumenshchev}, \&
  {Abramowicz}}]{cdaf}
{Narayan}, R., {Igumenshchev}, I.~V., \& {Abramowicz}, M.~A. 2000, \apj, 539,
  798

\bibitem[{{Narayan} {et~al.}(1997){Narayan}, {Kato}, \& {Honma}}]{nkh}
{Narayan}, R., {Kato}, S., \& {Honma}, F. 1997, \apj, 476, 49

\bibitem[{{Narayan} {et~al.}(1996){Narayan}, {McClintock}, \& {Yi}}]{nmyi96}
{Narayan}, R., {McClintock}, J.~E., \& {Yi}, I. 1996, \apj, 457, 821+

\bibitem[{{Narayan} {et~al.}(2002){Narayan}, {Quataert}, {Igumenshchev}, \&
  {Abramowicz}}]{nqia}
{Narayan}, R., {Quataert}, E., {Igumenshchev}, I.~V., \& {Abramowicz}, M.~A.
  2002, \apj, submitted (astro-ph/02033026), \

\bibitem[{{Narayan} \& {Yi}(1994)}]{ny94}
{Narayan}, R. \& {Yi}, I. 1994, \apjl, 428, L13

\bibitem[{{Narayan} \& {Yi}(1995{\natexlab{a}})}]{ny2d}
---. 1995{\natexlab{a}}, \apj, 444, 231

\bibitem[{{Narayan} \& {Yi}(1995{\natexlab{b}})}]{nyi}
---. 1995{\natexlab{b}}, \apj, 452, 710

\bibitem[{{Novikov} \& {Thorne}(1973)}]{nt}
{Novikov}, I.~D. \& {Thorne}, K.~S. 1973, in Black Holes, ed. C.~{DeWitt} \&
  B.~{DeWitt} (New York: Gordon \& Breach), 334

\bibitem[{{Ogilvie}(1999)}]{ogilvie}
{Ogilvie}, G.~I. 1999, \mnras, 306, L9

\bibitem[{{Osaki}(1974)}]{osaki2}
{Osaki}, Y. 1974, \pasj, 26, 429

\bibitem[{{Osaki}(1996)}]{osaki}
---. 1996, \pasp, 108, 39

\bibitem[{{Papaloizou} \& {Pringle}(1977)}]{pp77}
{Papaloizou}, J. \& {Pringle}, J.~E. 1977, \mnras, 181, 441

\bibitem[{{Peitz} \& {Appl}(1997)}]{pa97}
{Peitz}, J. \& {Appl}, S. 1997, \mnras, 286, 681

\bibitem[{{Piran}(1978)}]{piran}
{Piran}, T. 1978, \apj, 221, 652

\bibitem[{{Popham} \& {Gammie}(1998)}]{pg98}
{Popham}, R. \& {Gammie}, C.~F. 1998, \apj, 504, 419

\bibitem[{{Popham} \& {Sunyaev}(2001)}]{popham}
{Popham}, R. \& {Sunyaev}, R. 2001, \apj, 547, 355

\bibitem[{{Poutanen}(1998)}]{poutanen}
{Poutanen}, J. 1998, in {Theory of Black Hole Accretion Disks}, ed.
  M.~{Abramowicz}, G.~{Bjornsson}, \& J.~{Pringle} (Cambridge University
  Press), 100

\bibitem[{{Pringle}(1976)}]{pringle76}
{Pringle}, J.~E. 1976, \mnras, 177, 65

\bibitem[{{Pringle}(1981)}]{pringle}
---. 1981, \araa, 19, 137

\bibitem[{{Pringle} \& {Rees}(1972)}]{pr72}
{Pringle}, J.~E. \& {Rees}, M.~J. 1972, \aap, 21, 1

\bibitem[{{Pringle} {et~al.}(1973){Pringle}, {Rees}, \& {Pacholczyk}}]{prp}
{Pringle}, J.~E., {Rees}, M.~J., \& {Pacholczyk}, A.~G. 1973, \aap, 29, 179

\bibitem[{{Quataert}(1998)}]{beta0}
{Quataert}, E. 1998, \apj, 500, 978

\bibitem[{{Quataert} \& {Gruzinov}(1999)}]{beta1}
{Quataert}, E. \& {Gruzinov}, A. 1999, \apj, 520, 248

\bibitem[{{Quataert} \& {Gruzinov}(2000)}]{cdaf2}
---. 2000, \apj, 539, 809

\bibitem[{{Quataert} \& {Narayan}(1999)}]{ADIOSspec}
{Quataert}, E. \& {Narayan}, R. 1999, \apj, 520, 298

\bibitem[{{Rees} {et~al.}(1982){Rees}, {Phinney}, {Begelman}, \&
  {Blandford}}]{rees}
{Rees}, M.~J., {Phinney}, E.~S., {Begelman}, M.~C., \& {Blandford}, R.~D. 1982,
  \nat, 295, 17

\bibitem[{{Regev}(1991)}]{regev}
{Regev}, O. 1991, in Structure and Emission Properties of Accretion Disks, IAU
  Colloq. 129, ed. J.-P.~L. C.~Bertout, S. Collin-Souffrin (Editions
  Frontieres), 311

\bibitem[{{Shakura} \& {Sunyaev}(1973)}]{ss}
{Shakura}, N.~I. \& {Sunyaev}, R.~A. 1973, \aap, 24, 337

\bibitem[{{Shakura} \& {Sunyaev}(1976)}]{ss76}
---. 1976, \mnras, 175, 613

\bibitem[{{Shapiro} {et~al.}(1976){Shapiro}, {Lightman}, \& {Eardley}}]{sle}
{Shapiro}, S.~L., {Lightman}, A.~P., \& {Eardley}, D.~M. 1976, \apj, 204, 187

\bibitem[{{Shaviv}(2001)}]{shaviv}
{Shaviv}, N.~J. 2001, \mnras, 326, 126

\bibitem[{{Smak}(1983)}]{smak}
{Smak}, J. 1983, \apj, 272, 234

\bibitem[{{Smak}(1984)}]{smak84}
---. 1984, Acta Astronomica, 34, 161

\bibitem[{{Smak}(1994)}]{smak94}
---. 1994, Acta Astronomica, 44, 265

\bibitem[{{Smak}(2000)}]{smakox}
---. 2000, New Astronomy Review, 44, 171

\bibitem[{{Spruit}(1987)}]{spruit}
{Spruit}, H.~C. 1987, \aap, 184, 173

\bibitem[{{Spruit} {et~al.}(1987){Spruit}, {Matsuda}, {Inoue}, \&
  {Sawada}}]{spruitetal}
{Spruit}, H.~C., {Matsuda}, T., {Inoue}, M., \& {Sawada}, K. 1987, \mnras, 229,
  517

\bibitem[{{Stepinski} {et~al.}(1993){Stepinski}, {Reyes-Ruiz}, \&
  {Vanhala}}]{stepinski}
{Stepinski}, T.~F., {Reyes-Ruiz}, M., \& {Vanhala}, H.~A.~T. 1993, Icarus, 106,
  77

\bibitem[{{Stone} {et~al.}(1999){Stone}, {Pringle}, \& {Begelman}}]{spb1999}
{Stone}, J.~M., {Pringle}, J.~E., \& {Begelman}, M.~C. 1999, \mnras, 310, 1002

\bibitem[{{Sunyaev} \& {Shakura}(1975)}]{ss75}
{Sunyaev}, R.~A. \& {Shakura}, N.~I. 1975, Soviet Astronomy Letters, 1, 158

\bibitem[{{Szuszkiewicz} \& {Miller}(2001)}]{szu}
{Szuszkiewicz}, E. \& {Miller}, J.~C. 2001, \mnras, 328, 36

\bibitem[{{Terquem}(2001)}]{terquem}
{Terquem}, C. 2001, in Proceedings of the Aussois 2000 Summer School on Star
  formation and the Physics of Young Stars, ed. J.~{Bouvier} \& J.-P. {Zahn}
  (astro-ph/0107408)

\bibitem[{{Warner}(1995)}]{cvs}
{Warner}, B. 1995, {Cataclysmic variable stars} (Cambridge University Press)

\end{thebibliography}

\end{document}